\documentclass[aip,pop,reprint,numerical]{revtex4-1}
\usepackage{dcolumn}
\usepackage[english]{babel}
\usepackage{graphicx}
\usepackage{indentfirst}
\usepackage[utf8]{inputenc}
\usepackage{enumerate}
\usepackage{bm}
\usepackage{units}
\usepackage{color}
\usepackage{url}
\usepackage[hidelinks,colorlinks,breaklinks=true]{hyperref}
\hypersetup{
  colorlinks=true,    % false: boxed links; true: colored links
  linkcolor=blue,     % color of internal links
  citecolor=blue,     % color of links to bibliography
  filecolor=magenta,  % color of file links
  urlcolor=blue,
  bookmarksdepth=4
}

\begin{document}
\def\va{v_{\parallel}}
\def\ve{v_{\perp}}
\def\ua{u_{\parallel}}
\def\ue{u_{\perp}}
\def\ka{k_{\parallel}}
\def\ke{k_{\perp}}
\def\na{N_{\parallel}}

\title{Collisional damping rates for plasma waves}

\author{S. F. Tigik}
\email{sabrina.tigik@ufrgs.br}
\author{L. F. Ziebell}
\email{luiz.ziebell@ufrgs.br}
\affiliation{Instituto de F\'{\i}sica, Universidade Federal
do Rio Grande do Sul, 91501-970 Porto Alegre, RS, Brazil}
\author{P. H. Yoon}
\email{yoonp@umd.edu}
\affiliation{Institute for Physical Science \& Technology,
University of Maryland, College Park, MD 20742, USA}
\affiliation{School of Space Research, Kyung Hee University,
Yongin, Gyeonggi 446-701, South Korea}

\begin{abstract}
The distinction between the plasma dynamics dominated by
collisional transport versus collective processes has
never been rigorously addressed until recently.
A recent paper [Yoon et al., Phys. Rev. E {\bf 93}, 033203 (2016)]
formulates for the first time, a unified kinetic theory
in which collective processes and collisional dynamics
are systematically incorporated from first principles.
One of the outcomes of such a formalism is the rigorous
derivation of collisional damping rates for Langmuir and
ion-acoustic waves, which can be contrasted to the
heuristic customary approach. However, the results are
given only in formal mathematical expressions. The present
Brief Communication numerically evaluates the rigorous collisional
damping rates by considering the case of plasma particles
with Maxwellian velocity distribution function so as to
assess the consequence of the rigorous formalism
in a quantitative manner. Comparison with the heuristic
(``Spitzer'') formula shows that the accurate damping
rates are much lower in magnitude than the conventional
expression, which implies that the traditional approach
over-estimates the importance of attenuation of
plasma waves by collisional relaxation process.
Such a finding may have a wide applicability ranging
from laboratory to space and astrophysical plasmas.
\end{abstract}

%\pacs{52.35.-g, 52.35.Fp, 52.35.Mw}

%52.35.Ra Plasma turbulence
%52.35.Mw Nonlinear waves and nonlinear wave propagation
%52.35.Qz Micro instabilities(beam-plasma, two-stream...)
%52.35.Fp Electrostatic waves and oscillations
%52.35.-g Waves, oscillations, and instabilities
%         in plasmas and intense beams

\maketitle

In a recently published paper \cite{YZKS16}, the formalism
of plasma kinetic theory was revisited, and a set of coupled
equations were derived, which describe the dynamical evolution
of the velocity distribution functions of plasma particles and
the spectral wave energy densities associated with electrostatic
waves. Reference \cite{YZKS16} follows the standard
weak turbulence perturbative ordering, except that unlike the
textbook approaches, which take into account only the collective
eigenmodes in the linear and nonlinear wave-particle interactions,
the new formalism includes the effects of non-collective
fluctuations emitted by thermal particles. It is shown that
the non-collective fluctuations, which had been largely ignored
in the literature hitherto, are responsible for collisional
effects in both the particle and wave equations. Specifically,
Ref.\ \cite{YZKS16} demonstrates that the inclusion
of non-collective part of thermal fluctuations leads to the
collision integral, while the collective eigenmodes are
responsible for the usual quasi-linear diffusion (plus the
velocity friction) term(s) in the particle kinetic equation.
As for the collectively excited waves, which satisfy the
dispersion relations, and are thus eigenmodes, the influence
of non-collective thermal fluctuations rigorously lead to
the collisional wave damping of the collective waves, as well
as the emission of these waves by particle collisions (i.e.,
bremsstrahlung emission of electrostatic eigenmodes).
Such a derivation, without any {\it ad hoc} additions,
was done for the first time.

If one is interested only in the collisional relaxation for
collision-dominated plasmas, then transport processes can be
legitimately discussed solely on the basis of well-known
collisional kinetic equation \cite{Helander2002, Zank2014}.
Collisional transport is important for high density plasmas
such as in the solar interior. In the opposite limit,
if one's concern is only on relaxation processes that
involve collective oscillations and waves, then various
nonlinear theories of plasma turbulence may be employed
\cite{Sitenko1982, Melrose1986, Balescu2005, Treumann1997}. Collective
processes dominate rarefied plasmas, which characterize
most of the heliosphere, interstellar, and
intergalactic environments.

It is the dichotomy that separates the purely collisional
versus purely collective descriptions that had not been
rigorously bridged until the recent work \cite{YZKS16}.
There are intermediate situations where both collisional
and collective processes must be treated together, such as
the solar x ray bremsstrahlung radiation sources
\cite{McClements87b, HannahKS09, RatcliffeK14}, or in the
Earth's ionospheric plasma where collisional conductivity
becomes important \cite{Kelley2009} (note that for the Earth's
ionosphere, the dominant collisional process is the
charged particle collisions with the neutrals, however).
For such situations, there was a general lack of
satisfactory theories, which one may bring to bear in
order to address the necessary physics, until recently.
Instead, it had been a common practice to introduce
collisional damping in the wave evolution as an
indirect effect of assuming a collisional operator in the
particle equation, and define an effective collision
frequency \cite{McClements87b, HannahKS09, RatcliffeK14,
Comisar1963, WoodN69, Melrose80, Benz2002, Lifshitz1981}. However,
such a procedure is tantamount to inserting the collisional
dissipation by hand, as it were, to the governing
microscopic equation which describes fundamentally
collision-free situation. Consequently, strictly speaking,
the method is at best, heuristic. Nevertheless, such an
{\it ad hoc} prescription is widely practiced in the
plasma physics literature.

Thus, in the literature, often a governing equation
is adopted,
%(1)
\begin{equation}
\left[\frac{\partial}{\partial t}+{\bf v}\cdot
\nabla+e_a\left({\bf E}+\frac{\bf v}{c}
\times{\bf B}\right)\cdot\frac{\partial}{\partial{\bf p}}
\right]f_a=C_a(f_a),
\label{1}
\end{equation}
where $C_a(f_a)$ represents the collision integral,
$a$ denoting particle species ($a=e$ for electrons,
$a=i$ for ions). If $f_a$, ${\bf E}$ and ${\bf B}$ in
the above represent the averaged one-particle distribution
function and average fields, then Eq.\ (\ref{1}) represents
the correct collisional kinetic equation \cite{Helander2002, Zank2014}.
However, if these represent the total (average plus fluctuation),
then they become microscopic one-particle distribution
function and microscopic fields. For such a case, the
right-hand side of Eq.\ (\ref{1}) should be zero, since
microscopic equations are reversible. As shown in Ref.\
\cite{YZKS16}, the irreversibility (signified
by collision operator on the right-hand side) enters the
problem only as a result of statistical averages and the
loss of information. Nonetheless, the standard procedure
in the literature is to interpret $f_a$ and field vectors
as microscopic quantities, and employ expansion for
small-amplitude perturbations. Upon replacing the collision
operator by an effective collision frequency,
$C_a(f_a)\approx-\nu_{\rm coll}f_a$, the effective
particle collision frequency is absorbed into the
wave-particle resonance condition, and ends up
as part of the imaginary part of the wave frequency,
corresponding to a damping effect on the waves.
As a consequence of the above-described recipe,
known as the ``Spitzer approximation'' in the literature,
one may obtain the collisional damping rate
for Langmuir waves, given by
%(2)
\begin{equation}
\gamma_{\rm coll}=-\frac{\pi n_ee^4\ln\Lambda}{m_e^2v_{Te}^3},
\label{2}
\end{equation}
where $v_{Te}=(2T_e/m_e)^{1/2}$ is the electron thermal
speed and $\Lambda=\lambda_{De}T_e/e^2=4\pi n_e\lambda_{De}^3$
is a constant. Note that $\Lambda$ represents the total
number of electrons in a sphere whose radius is equal to
the Debye length, $\lambda_{De}=[T_e/(4\pi n_ee^2)]^{1/2}$.
Here, $m_e$, $T_e$, and $n_e$ stand for electron mass,
electron temperature (in the unit of energy), and
electron density, respectively. Note that Eq.\ (\ref{2})
implies that the collisional damping rate is constant
and does not depend on wave vector (or wave frequency).

Reference \cite{YZKS16}, in contrast, shows that
the accurate collisional damping rates for plasma waves,
that is, Langmuir ($L$) and ion-acoustic ($S$) waves, are
much more complicated that is indicated by the approximate
formula (\ref{2}) in that the correct formulae exhibit
dependence on wave number (and thus, frequency).
However, the final results were given only in terms of
formal expressions so that it is difficult to assess
the consequence of the new formulation. The purpose
of the present Brief Communication is to carry out numerical analysis
so that one may understand the significance, or
lack thereof, of the new findings in a quantitative way.

We start with the formal and rigorous expression for the
collisional damping rates for $L$ and $S$ waves, as given
by Eq.\ (4.44) in Ref.\ \cite{YZKS16}
%(3,4)
\begin{eqnarray}
\gamma_{\bf k}^{L({\rm coll})} &=&
\omega_{\bf k}^L\frac{4n_ee^4\omega_{pe}^2}
{T_e^2}\int d{\bf k}'\frac{({\bf k}\cdot{\bf k'})^2
\lambda_{De}^4}{k^2{k'}^4|\epsilon({\bf k'},
\omega_{\bf k}^L)|^2}
\nonumber\\
&& \times\left(1+\frac{T_e}{T_i}
+({\bf k}-{\bf k'})^2\lambda_{De}^2\right)^{-2}
\label{3}\\
&& \times \int d{\bf v}\,{\bf k'}\cdot
\frac{\partial F_e({\bf v})}{\partial {\bf v}}
\delta(\omega_{\bf k}^L-{\bf k'}\cdot{\bf v}),
\nonumber\\
\gamma_{\bf k}^{S({\rm coll})} &=&
\mu_{\bf k}\omega_{\bf k}^L
\frac{n_ee^4\omega_{pe}^2}{T_e^2}\int d{\bf k}'
\frac{1}{k^2{k'}^4|\epsilon({\bf k'},\omega_{\bf k}^S)|^2}
\nonumber\\
&& \times\left(1+\frac{T_e}{T_i}
+({\bf k}-{\bf k'})^2\lambda_{De}^2\right)^{-2}
\label{4}\\
&& \times\left(1+\frac{2T_e}{T_i}
\frac{{\bf k}\cdot{\bf k}'}{k^2}\right)
\int d{\bf v}\,{\bf k'}\cdot
\frac{\partial}{\partial {\bf v}}
\nonumber\\
&& \times\left(F_e({\bf v})+\frac{m_e}{m_i}
F_i({\bf v})\right)\delta(\omega_{\bf k}^S
-{\bf k'}\cdot{\bf v}).
\nonumber
\end{eqnarray}
In the above, $\omega_{\bf k}^L=\omega_{pe}\left(
1+\frac{3}{2}k^2\lambda_{De}^2\right)$ and
$\omega_{\bf k}^S=\omega_{pe}k\lambda_{De}
\sqrt{\frac{m_e}{m_i}\frac{1+3T_i/T_e}
{1+k^2\lambda_{De}^2}}$
designate Langmuir and ion-sound mode dispersion
relations, respectively, $m_i$ and $T_i$ being the
ion (proton) mass and temperature, respectively,
and $\omega_{pe}=(4\pi n_ee^2/m_e)^{1/2}$ is the
plasma frequency. The ensemble-averaged one-particle
distribution function $F_a({\bf v})$ is normalized
to unity, $\int d{\bf v}F_a({\bf v})=1$. The quantity
$\mu_{\bf k}$ is defined by $\mu_{\bf k}=k^3\lambda_{De}^3
\sqrt{\frac{m_e}{m_i}\left(1+\frac{3T_i}{T_e}\right)}$,
and $\epsilon({\bf k},\omega_{\bf k}^L)$
and $\epsilon({\bf k},\omega_{\bf k}^S)$ are the
dielectric constants,
\begin{displaymath}
\epsilon({\bf k},\omega)=1+\sum_a\frac{\omega_{pa}^2}{k^2}
\int d{\bf v}\frac{{\bf k}\cdot\partial F_a/\partial{\bf v}}
{\omega-{\bf k}\cdot{\bf v}+i0}.
\end{displaymath}
Evidently, Eqs.\ (\ref{3}) and (\ref{4}) are far more
sophisticated than the simple expression (\ref{2}).
The question is what is the actual implication of these
results? Specifically, to what extent does the approximation
(\ref{2}) conform with the rigorous results
(\ref{3}) and (\ref{4}), and if not, what is
the extent of the discrepancy?

In order to quantitatively analyze Eqs.\ (\ref{3}) and
(\ref{4}), it is advantageous to introduce suitable
dimensionless quantities,
%(5)
\begin{equation}
{\bf u}=\frac{\bf v}{v_{Te}},\quad
z=\frac{\omega}{\omega_{pe}},\quad
{\bf q}=\frac{{\bf k}v_{Te}}{\omega_{pe}}
={\bf k}\sqrt{2}\lambda_{De},
\label{5}
\end{equation}
and rewrite the collisional damping rates (\ref{3}) and
(\ref{4}) in normalized form,
%(6,7)
\begin{eqnarray}
\gamma_{\bf q}^{L({\rm coll})} &\equiv& 
\frac{\gamma_{\bf k}^{L({\rm coll})}}{\omega_{pe}}
= \frac{2gz_{\bf q}^L}{q^2}\int d{\bf q}'
\frac{({\bf q}\cdot {\bf q'})^2}{{q'}^4 
|\epsilon({\bf q'},z_{\bf q}^L)|^2}
\nonumber\\
&& \times\left(1+\frac{T_e}{T_i}
+\frac{({\bf q}-{\bf q'})^2}{2}\right)^{-2}
\label{6}\\
&& \times \int d{\bf u}\,{\bf q'}\cdot
\frac{\partial \Phi_e({\bf u})}{\partial {\bf u}}
\delta(z_{\bf q}^L-{\bf q'}\cdot{\bf u}),
\nonumber\\
\gamma_{\bf q}^{S ({\rm coll})} &\equiv& 
\frac{\gamma_{\bf q}^{S({\rm coll})}}{\omega_{pe}}
=\frac{2gz_{\bf q}^L}{q^2}\int\frac{d{\bf q}'}
{{q'}^4|\epsilon({\bf q'},z_{\bf q}^S)|^2}
\nonumber\\
&& \times\left(1+\frac{T_e}{T_i}+
\frac{({\bf q}-{\bf q'})^2}{2}\right)^{-2}
\nonumber\\
&& \times\left(1+\frac{2T_e}{T_i}
\frac{{\bf q}\cdot{\bf q}'}{q^2}\right)
\int d{\bf u}\, {\bf q'}\cdot\frac{\partial}{\partial {\bf u}}
\label{7}\\
&&\times\left(\Phi_e({\bf u})+\frac{m_e}{m_i}
\Phi_i({\bf u})\right)\delta(z_{\bf q}^S
-{\bf q'}\cdot{\bf u}),
\nonumber
\end{eqnarray}
where in dimensionless form, the dispersion relations
are given by $z_{\bf q}^L=1+\frac{3q^2}{4}$ and
$z_{\bf q}^S=q\sqrt{\frac{m_e}{m_i}\frac{1+3T_i/T_e}{2+q^2}}$.
In Eqs.\ (\ref{6}) and (\ref{7}), the quantity $g$
is defined by
%(8)
\begin{equation}
g=\frac{1}{2^{3/2}(4\pi)^2{n}_e\lambda_{De}^3}
=\frac{1}{2^{3/2}(4\pi\Lambda)},
\label{8}
\end{equation}
which is related to the parameter $\Lambda$ discussed
earlier. The quantity $g$ is an effective
``{\it plasma parameter}'' in that it is related
to the inverse of the number of particles in a ``Debye
sphere.''

Let us assume that ions and electrons have isotropic
Maxwellian velocity distributions,
%(9)
\begin{eqnarray}
\Phi_a({\bf u}) &=& v_e^3F_a({\bf v})
=\frac{1}{\pi^{3/2}}\left(\frac{m_a}{m_e}
\frac{T_e}{T_a}\right)^{3/2}
\nonumber\\
&& \times\exp\left(-\frac{m_a}{m_e}
\frac{T_e}{T_a}u^2\right).
\label{9}
\end{eqnarray}
Then the dielectric constants appearing in the
denominators of Eqs.\ (\ref{3}) and (\ref{4}) are
given by the following:
%(10,11)
\begin{eqnarray}
\epsilon({\bf q}',z_{\bf q}^L) &=&
1+\frac{2}{{q'}^2}\left[1+\frac{z_{\bf q}^L}{q'}
Z\left(\frac{z_{\bf q}^L}{q'}\right)\right],
\label{10}\\
\epsilon({\bf q}',z_{\bf q}^S) &=&
1+\frac{2}{{q'}^2}\left[1+\frac{z_{\bf q}^S}{q'}
Z\left(\frac{z_{\bf q}^S}{q'}\right)\right]
\nonumber\\
&& +\frac{T_e}{T_i}\frac{2}{{q'}^2}\left\{1+
\left(\frac{m_i}{m_e}\frac{T_e}{T_i}\right)^{1/2}
\frac{z_{\bf q}^S}{q'}\right.
\nonumber\\
&& \left.\times Z\left[\left(\frac{m_i}{m_e}
\frac{T_e}{T_i}\right)^{1/2}
\frac{z_{\bf q}^S}{q'}\right]\right\}.
\label{11}
\end{eqnarray}

For Maxwellian velocity distribution (\ref{9}),
the velocity integral $\int d{\bf u}$ in Eqs.\ (\ref{6})
and (\ref{7}) can be carried out analytically
upon making use of the resonance delta function
conditions. One may also perform the angular
integration associated with the ${\bf q}'$ vector
integral, which reduces Eqs.\ (\ref{6}) and (\ref{7})
in the form that involves a single $q'$ integration,
%(12,13)
\begin{eqnarray}
\gamma_{\bf q}^{L({\rm coll})} &=& -(16\pi^{1/2}g) 
\frac{(z_{\bf q}^L)^2}{q^2}\int_0^\infty\frac{dq'}
{|\epsilon(q',z_{\bf q}^L)|^2}
\nonumber\\
&& \times\left(\frac{2B^2-A^2}{B^2-A^2}
-\frac{B}{A}\ln\frac{B+A}{B-A}\right)
\nonumber\\
&& \times\frac{1}{{q'}^3}
\exp\left(-\frac{(z_{\bf q}^L)^2}{{q'}^2}\right),
\label{12}\\
\gamma_{\bf q}^{S({\rm coll})} &=& -(16\pi^{1/2}g)
\frac{\mu_{\bf q}z_{\bf q}^Lz_{\bf q}^S}{q^2}
\int_0^\infty\frac{dq'}{|\epsilon(q',z_{\bf q}^S)|^2}
\nonumber\\
&& \times\left[\frac{4}{B^2-A^2}-\frac{T_e}{T_i}
\frac{q'}{q}\frac{1}{q^2{q'}^2}\right.
\nonumber\\
&& \left.\times\left(\frac{2AB}{B^2-A^2}
-\ln\frac{B+A}{B-A}\right)\right]
\nonumber\\
&& \times\sum_{a=e,i}\frac{T_e}{T_a}\left(
\frac{m_a}{m_e}\frac{T_e}{T_a}\right)^{1/2}
\nonumber\\
&& \times\frac{1}{{q'}^3}\exp\left(-\frac{m_a}{m_e}
\frac{T_e}{T_a}\frac{(z_{\bf q}^S)^2}{q'^2}\right),
\label{13}
\end{eqnarray}
where we have defined
%(14)
\begin{eqnarray}
A &=& -2qq',
\nonumber\\
B &=& 2\left(1+\frac{T_e}{T_i}\right)+q^2+{q'}^2.
\label{14}
\end{eqnarray}
For reference, the customary heuristic collisional damping
rate (\ref{2}), derived under the ``Spitzer approximation,''
which is applicable for Langmuir wave, is given in
normalized form by
%(15)
\begin{eqnarray}
\bar{\gamma}_{\rm coll} &\equiv&
\frac{\gamma_{\rm coll}}{\omega_{pe}}
=-\frac{\pi n_ee^4\ln\Lambda}{m_e^2v_{Te}^3\omega_{pe}}
\nonumber\\
&=& -\pi g\ln\left(\frac{1}{2^{3/2}(4\pi g)}\right).
\label{15}
\end{eqnarray}

For comparison, we also discuss the collisionless damping,
also known as Landau damping, which is well-known.
From Eq.\ (3.24) of Ref.\ \cite{YZKS16}, we have
the Landau damping rates for $L$ and $S$ waves,
%(16)
\begin{eqnarray}
\gamma_{\bf k}^L &=& \frac{\pi\omega_{\bf k}^L
\omega_{pe}^2}{2k^2}\int d{\bf v}\,{\bf k}\cdot
\frac{\partial F_e({\bf v})}{\partial{\bf v}}
\delta(\omega_{\bf k}^L-{\bf k}\cdot{\bf v}),
\nonumber\\
\gamma_{\bf k}^S &=& \frac{\pi\mu_{\bf k}\omega_{\bf k}^L
\omega_{pe}^2}{2k^2}\int d{\bf v}\,{\bf k}\cdot
\frac{\partial}{\partial {\bf v}}
\nonumber\\
&& \times\left(F_e({\bf v})
+\frac{m_e}{m_i}F_i({\bf v})\right)
\delta(\omega_{\bf k}^S-{\bf k}\cdot{\bf v}),
\label{16}
\end{eqnarray}
which are textbook results. Making use of dimensionless
variables, the above expressions are rewritten as
%(17)
\begin{eqnarray}
\gamma_{\bf q}^L &=& -\frac{\pi^{1/2}(z_{\bf q}^L)^2}{q^3}
\exp\left(-\frac{(z_{\bf q}^L)^2}{q^2}\right),
\nonumber\\
\gamma_{\bf q}^S &=& -\frac{\pi^{1/2}\mu_{\bf q}
z_{\bf q}^Lz_{\bf q}^S}{q^3}\sum_{a=e,i}\frac{T_e}{T_a}
\left(\frac{m_a}{m_e}\frac{T_e}{T_a}\right)^{1/2}
\nonumber\\
&& \times\exp\left(-\frac{m_a}{m_e}
\frac{T_e}{T_a}\frac{(z_{\bf q}^S)^2}{q^2}\right).
\label{17}
\end{eqnarray}

\begin{figure}[h]
\centering
\includegraphics[width=\columnwidth]{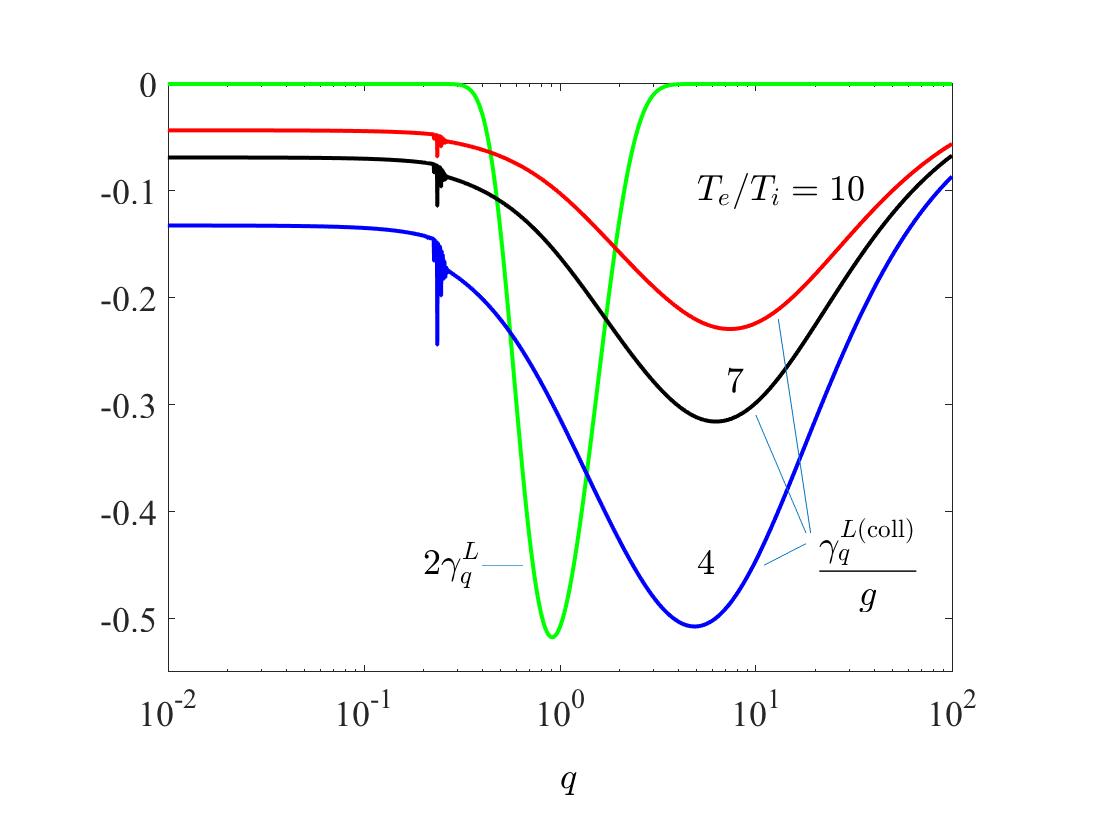}
\caption{Normalized collisional damping for $L$ waves,
$\gamma_q^{L({\rm coll})}/g$, vs normalized wavenumber $q$,
for three values of the ratio $T_e/T_i$.
The dimensionless Landau damping rate $\gamma_q^L$
is also plotted in green. Note that the Landau
damping rate is {\it not} divided by the factor
$g$. The factor 2, which multiplies $\gamma_q^L$
is for the sake of visual presentation.}
\label{fig1}
\includegraphics[width=\columnwidth]{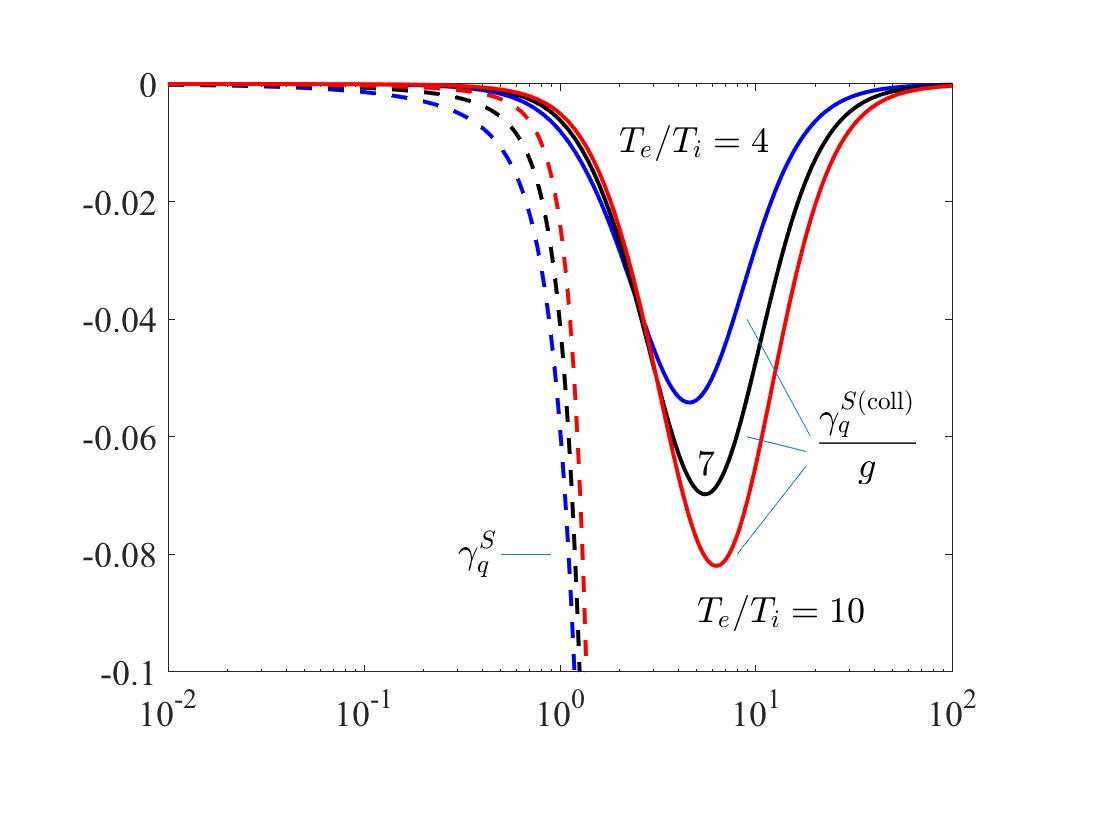}
\caption{Normalized collisional damping for $S$ waves,
$\gamma_q^{S({\rm coll})}/g$, vs normalized wavenumber $q$,
for three values of the ratio $T_e/T_i$.}
\label{fig2}
\end{figure}

In Fig.\ \ref{fig1} we plot the normalized collisional
$L$ mode damping rate divided by $g$,
$\gamma_q^{L({\rm coll})}/g$, as a function of
dimensionless wave number $q$, for three values of
the temperature ratio $T_e/T_i= 10$ (red),
7 (black), and 4 (blue). It is seen that for the range
of temperature ratios considered the damping rate is
maximum for $q$ between 5 and 9, approximately, and
that the growth rate increases with decreasing
temperature ratio $T_e/T_i$, for the entire range of
wavelengths. In contrast, the approximate collisional
damping rate divided by $g$, $\bar{\gamma}_{\rm coll}/g
=\pi\ln\left[2^{3/2}(4\pi g)\right]$, is independent of
the normalized wave number $q$, but the result depends
on $g$. In general, the plasma parameter $g$ must be
small by definition, so we consider several different
choices, $g=10^{-10}$, $10^{-8}$, $10^{-6}$, and $10^{-4}$.
For these choices, we find that $\bar{\gamma}_{\rm coll}/g
\sim-61.12$, $-46.6524$, $-32.1849$, and $-17.7173$,
which are all far higher in absolute value than
those depicted in Fig.\ \ref{fig1}. This shows that
the use of incorrect collisional damping rate may
greatly over-estimate the actual damping rate.

We also superpose in Fig.\ \ref{fig1}, the collisionless
(Landau) damping rate for Langmuir wave [i.e., the
first equation in (\ref{17})] vs $q$ (green). We multiplied the
damping rate by factor 2 for visual reason. Note that
the Landau damping rate is {\it not} divided by the
plasma parameter $g$, so that the actual magnitude
of the ``collisionless'' damping rate will greatly
exceed that of the ``collisional'' damping rate by
factor $1/g\gg1$. This shows that over the range of
wave numbers over which the most important linear and
nonlinear wave-particle interactions are expected to
take place, the collisional damping of the Langmuir
wave will be practically ignorable. However,
it is interesting to note that for small wave number
domain ($q\ll1$) for which the Landau damping rate
becomes negligible, the collisional damping rate
remains finite. In the collisionless plasmas
the undamped Langmuir waves in the long wavelength
regime are supposed to lead to the so-called condensation
phenomenon, where the wave energy accumulates without
undergoing Landau damping. Over a long time period,
the Zakharov strong turbulence effect is supposed to
come into play in order to dissipate the accumulated
wave energy \cite{Zakharov1972}. However, the
present finding suggests that the collisional damping
may contribute to the dissipation of the Langmuir wave
energy in such a wavelength regime. We also not
that for large $q$, the collisionless (Landau)
damping rate eventually becomes exponentially
weak. In contrast, the collisional damping rate
may overcome the Landau damping rate, which makes
sense, since for extremely short wavelength
the binary collision may lead to the damping of
plasma waves.

Before we close and for the sake of completeness,
we plot in Fig.\ \ref{fig2} the normalized
collisional damping for $S$ waves
divided by $g$, $\gamma_q^{S({\rm coll})}/g$,
as a function of wave number $q$, for the same
three values of the temperature ratio considered
in Fig.\ \ref{fig1}, that is, $T_e/T_i= 4$, 7, and 10.
The same color scheme is used to indicate the three cases.
Unlike the case of $L$ mode, the collisional damping
rate for $S$ mode does not asymptotically approach
a finite value for $q\to0$. We also superpose the
collisionless (Landau) damping rates for $S$ waves vs $q$,
but since $\gamma_q^S$ depends on $T_e/T_i$, we use the
same color scheme to indicate the three difference
choices of $T_e/T_i$, except that we plot the
collisionless damping rate with dashes. Again, we note
that $\gamma_q^S$ is {\it not} divided by $g$, so that the
actual damping rate is much higher in magnitude than the
collisional damping rate $\gamma_q^{S({\rm coll})}$.
In the case of $S$ mode, it becomes evident that the
collisional damping plays no significant role whatsoever
when compared against the collisionless damping, and thus
the dynamical role of collisions on the dissipation of
ion-sound mode damping becomes totally negligible.

% \begin{figure}[h]
% \centering
% \includegraphics[width=\columnwidth]{F2}
% \caption{Normalized collisional damping for $S$ waves,
% $\gamma_q^{S({\rm coll})}/g$, vs normalized wavenumber $q$,
% for three values of the ratio $T_e/T_i$.}
% \label{fig2}
% \end{figure}

In the present Brief Communication we have investigated the formal
collisional damping rates derived in Ref.\ \cite{YZKS16},
by numerical means. It is found that the collisional damping
rates for Langmuir and ion-acoustic waves are much smaller
than the conventional expressions, which means that the
collisional damping has been over-estimated in the literature.
While the collisional damping for ion-sound wave is
totally negligible, the same for Langmuir wave becomes
finite, albeit small, in the region of infinite wave length
regime where collisionless Landau damping rate vanishes.
Such a property may potentially provide the necessary
dissipation mechanism in order to prevent the unchecked
accumulation of wave energy for long wavelength regime,
known as the Langmuir condensation problem.

The importance of the present work is quite obvious.
There are many physical situations where collisional
and collective effects are both important, both in
laboratory and space applications. The present analysis
is based upon the recent work \cite{YZKS16}, which
makes a simplifying assumption of electrostatic
interaction in field-free plasmas. For more realistic
applications electromagnetic interaction in magnetized
plasmas must be considered within the framework of
the collisional weak turbulence theory. Reference
\cite{YZKS16} and the present work may represent
the beginning of a new research paradigm.

This work has been partially supported by the Brazilian
agencies CNPq and FAPERGS.
P.H.Y. acknowledges support by NSF grants AGS1147759
and AGS1550566 to the University of Maryland.
The research at Kyung Hee University was supported by the
BK21-Plus grant from the National Research Foundation (NRF), Korea.
P.H.Y. also acknowledges the Science Award Grant from the GFT
Foundation to the University of Maryland.

\bibliographystyle{aipnum4-1}
\bibliography{refs}

\end{document}